\documentclass[jkps,preprint,fleqn,showpacs,showkeys]{revtex4}
\usepackage{graphicx}
\usepackage{amssymb}
\usepackage{amsmath}
\usepackage{bm}

\begin{document}
\setcounter{page}{0}
\title[]{
    {\tt $\nu$Oscillation}: a software package for
    computation and simulation of neutrino 
    oscillation and detection\footnote{The source code is available at {\tt https://github.com/railroad2/vOscillation}}
}

\author{Seunghyeok Jang} 
\author{Eunil Won} 
\author{Kyungmin Lee}
\email{kmlee@hep.korea.ac.kr}

\affiliation{Department of Physics, Korea University, Seoul 02841, Korea}

\author{Eunju Jeon} 
\author{Young Ju Ko} 
\email{yjko@ibs.re.kr}

\affiliation{Center for Underground Physics, Institute for Basic Science, Daejeon, 34048, Korea} 

\date[]{}

\begin{abstract}
The design of neutrino experiments and the analysis of neutrino data rely on precise computations of neutrino oscillations and scattering processes in general. Motivated by this,
we developed a unified software package that calculates the expected number and energy spectrum of neutrino events in the liquid scintillation detector taking into account the neutrino flux at production, the oscillations of neutrinos during propagation and their interactions in the detectors. 
We also implemented the calculation of neutrino flux from nuclear reactors, the Sun, and radioactive isotopes 
to explorer various experimental setups using a single package.
This software package is validated by reproducing the result
of calculations and observations in other publications. 
We also demonstrate the feasibility of this package by calculating
the sensitivity of a liquid scintillation detector, currently in planning,
to the sterile neutrinos.
This work is expected to be utilized to identify the physics potential
and optimize the design of future neutrino experiments.
\end{abstract}

\pacs{14.60.Pq,  13.15.+g, 29.40.Mc}

\keywords{neutrino oscillation, neutrino detector, detector simulation}

\maketitle

\section{Introduction}
 
The standard model (SM) of particle physics\cite{ref:SM, ref:SM2, ref:SM3, ref:SM4, ref:QCD, ref:qcd2}
describing
the electroweak and strong nuclear
interactions of subatomic particles in microscopic scales
has been validated extremely successfully for nearly half a century. 
However, there is one notable
deviation in the neutrino sector. 
The SM predicts
that neutrinos are precisely massless, yet experiments: 
Super-Kamiokande\cite{ref:superK_atmos}, SNO\cite{ref:sno2001, ref:SNO2}, and KamLAND\cite{ref:kamland} indicate
that neutrinos have small but finite masses, allowing quantum
oscillations among weak eigenstates.
To address this fundamental problem 
in modern particle physics, 
significant efforts are being directed toward 
ongoing and future neutrino experiments\cite{ref:hyperkamiokande,ref:dune}.
Those experimental efforts often require computation
of neutrino production and propagation
at different experimental conditions such as nuclear reactors, solar,  
accelerator, or radioactive source-based environments. 

For the detailed study of the phenomena with neutrinos, there have
been number of software packages developed to calculate 
the oscillation and detection of
neutrinos\cite{ref:n_packages, ref:n_packages2}.
A large
number of existing packages are theoretically oriented,
mathematically very rigorous,
and focused on the calculation of given Hamiltonian, including
non-standard interactions. 
For feasibility study
and design optimization of future experiments, 
a software tool that also offers enough precision and flexibility is desired
to carry out detailed studies on future neutrino experiments.

Motivated by these, we developed a unified software package, the
$\nu${\tt Oscillation}, which calculates 
the expected number and energy of neutrino events in the detector, taking into account the oscillations of neutrinos and their interactions in the liquid scintillation detectors.
This work is expected to be utilized to identify the physics potential
and optimize the design of future neutrino experiments.

In section II, we introduce the software implementation;
we present the application of our package to the reactor experiment in section III, 
solar neutrino experiment in section IV, 
and sterile neutrino experiment in section V;
and in section VI, we present our conclusion.

\section{The Software package}

The relation between flavor and mass eigenstates is
described by the lepton mixing matrix, known as the 
Pontecorvo–Maki–Nakagawa–Sakata (PMNS) matrix\cite{ref:pmns}
\begin{eqnarray}
|\nu_{\alpha} \rangle = \sum_i U^*_{\alpha i} |\nu_i \rangle,
\label{eq:PMNS}
\end{eqnarray}
where the flavor index $\alpha$ is $e,\mu$ or $\tau$,
the mass index $i$ runs over $1,2,3$, and $U_{\alpha i}$ are elements of the PMNS matrix. 
The time evolution of Eq. ~(\ref{eq:PMNS}) can be described by the time-dependent Schr\"odinger equation, which naturally involves the mixing of flavor eigenstates as the system evolves. This phenomenon is known as neutrino oscillation. 
The probability amplitude for the
neutrino of flavor $\alpha$ to oscillate into a different flavor $\beta$ 
at the time $t$ is given by $\langle \nu_\beta | \nu_\alpha(t) \rangle$, providing
an oscillation probability
\begin{eqnarray}
P(\nu_\alpha \rightarrow \nu_\beta) = | \langle \nu_\beta | \nu_\alpha(t) \rangle |^2
= \Bigg|
\sum_i U_{\beta i}U^*_{\alpha i} e^{ -i E_i t} 
\Bigg|^2,
\label{eq:osc_flavor}
\end{eqnarray}
where $E_i$ is the energy of the $i$-th mass eigenstate. 
Our package
computes Eq.~(\ref{eq:osc_flavor}) at the ultra-relativistic limit of neutrinos.
For the computation we use the analytic expressions of 
$P(\nu_\alpha \rightarrow \nu_\beta)$ found in Ref.\cite{ref:Zito} and the parameters in Particle Data Group 2014\cite{ref:pars_PDG2014}. 
Please note that the outdated parameters are used 
for comparison with other results to validate the software.

Our software package, {\tt $\nu$Oscillation}, 
mainly computes the time evolution of Eq.~(\ref{eq:osc_flavor}) 
from the initial neutrino flavors set by the users. 
The {\tt $\nu$Oscillation} also contains computation of 
inverse beta decay (IBD) 
cross section, neutrino-electron scattering
cross section, and the flux of
anti-neutrinos from nuclear reactors, and neutrinos from the Sun,
in order to be flexible to different
experimental scenarios. 

We also include one type of fictitious 
sterile neutrino in the computation of time-evolution of neutrinos
as an option to study the sensitivity of the
sterile neutrino search on proposed experiments.
In incorporate the sterile neutrino, the (3+1) model is implemented, which describes the neutrino mixing by a $4\times 4$ matrix. The parameters for sterile neutrino are sourced from Ref.  \cite{ref:sterile_parameters}. This can be generalized by involving more than one sterile neutrino, such as the (3+2) model, which can be implemented in the future releases.

Our code is written in {\tt C++} implemented as
{\tt C++} header and source codes utilizing libraries in 
{\tt ROOT}\cite{ref:root}, one of the open-source data analysis frameworks
actively used in the field of particle physics. Note also that the
{\tt ROOT} is the de facto standard analysis framework for the experimental
high energy physics community.
Intended users
of {\tt $\nu$Oscillation} are required to install {\tt ROOT} along with
{\tt $\nu$Oscillation}, but nothing more.  This allows users to utilize
{\tt $\nu$Oscillation} on any operating system that supports {\tt ROOT}. 
The source codes are available at
{\tt https://github.com/railroad2/vOscillation}.

\begin{figure}[htbp]
\includegraphics[width=\linewidth]{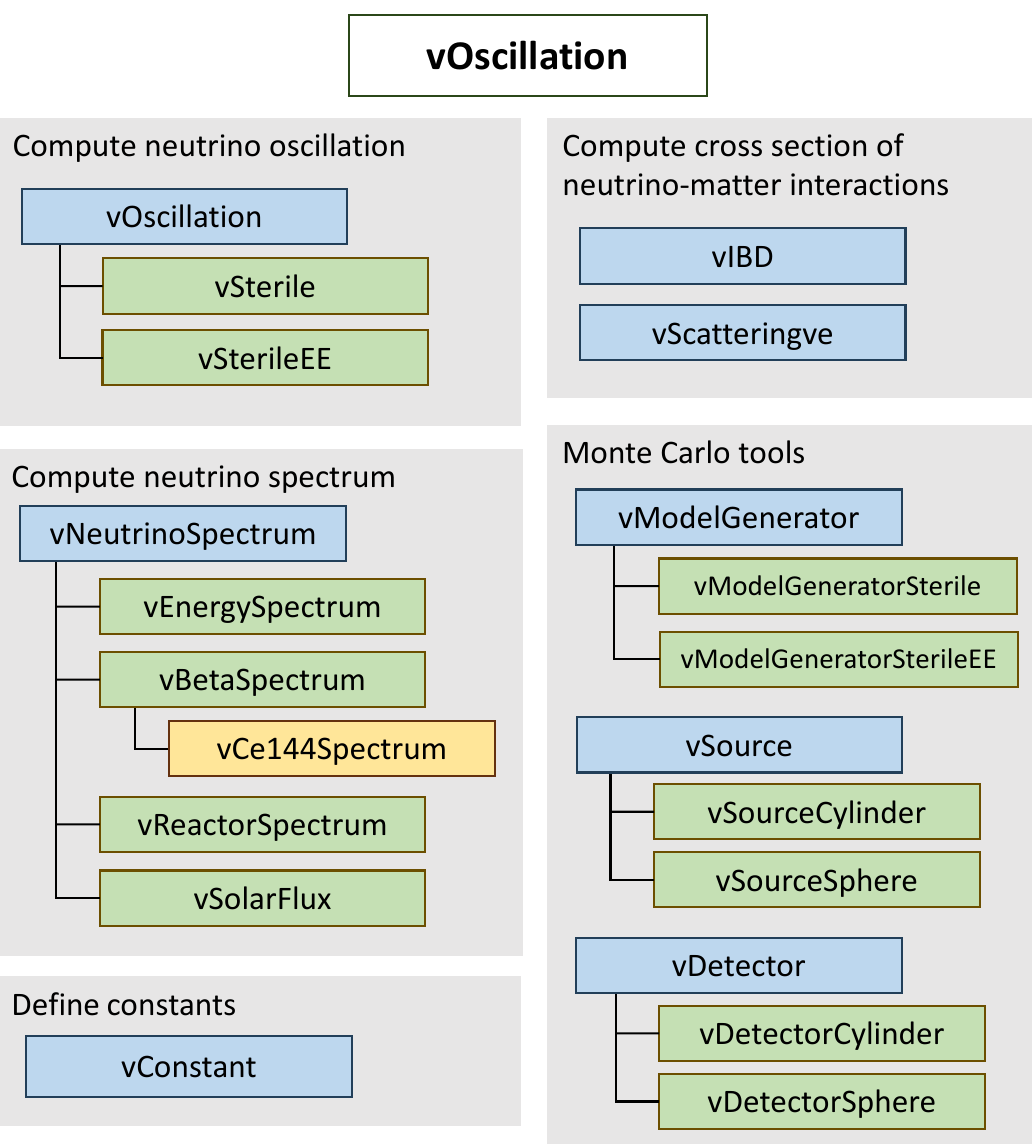}
\caption{
    Simplified functional
    block diagram of the $\nu${\tt Oscillation} package. This package is designed to calculate the expected numbers and energy distributions of neutrino events in the detectors for various experimental setups.
    It includes the classes to calculate the neutrino flux from nuclear reactors, the Sun, and radioactive sources. Neutrino oscillation during propagation is implemented for both traditional three-flavor case and (3+1) model, which includes a sterile neutrino.
    The package also implements calculations of the cross sections of IBD and neutrino-electron scattering to simulate interactions with detectors.
}
\label{fig:diagram}
\end{figure}

The {\tt $\nu$Oscillation} package has the 
following structure as shown in Fig. \ref{fig:diagram}.
The {\tt vOscillation} class has functions and variables to compute neutrino oscillations probabilities.   
Its subclass, {\tt vSterile}, introduces one type of sterile neutrino in 
the oscillation based on the (3+1) model. 
Note that the {\tt vSterileEE} class is a simplified implementation of the (3+1) model 
accounting only the electron flavor initial state.
Users can freely set the neutrino mixing parameters such as
$\Delta m^2_{ji}$ and $\theta_{ij}$ where $(i,j)$ can be
$(1,2), (1,3)$ or $(2,3)$. Here, $\Delta m^2_{ji}$ and $\theta_{ij}$ 
represent the mass squared difference between two-mass eigenstates of
a given neutrino pair and the mixing angle between them, respectively\cite{ref:Zito}. 
Users can also set charge-conjugation and parity $(CP)$ phase. 
The matter effect of the earth is not implemented in the current version of the package as it is not significant for our current applications with neutrino propagation distance of much less than 1000 km.

The calculation of energy spectra of neutrinos from radioactive sources, nuclear reactors, and the Sun is implemented in the three distinct classes. 
The {\tt vBetaSpectrum} class computes the beta spectrum of radioactive sources, employing the formula derived by Fermi\cite{ref:Fermi_Wilson}. Its subclass, {\tt vCe144Spectrum}, specifies the mass number, released energy, and atomic number of the daughter nucleus to compute the beta spectrum from ${}^{144}\text{Ce}$.
In a similar manner, the computation of energy spectra for various radioactive sources can be implemented.
The {\tt vReactorSpectrum} and {\tt vSolarFlux} classes are responsible for computing the neutrino spectra from the reactors and the Sun.
The details of these implementations will be described in the following sections.

The {\tt vIBD} class computes the IBD cross section using the formula with first-order corrections 
for recoil and weak magnetism\cite{ref:Vogel}. 
The {\tt vScatteringve} class computes the scattering cross section between neutrino and electron based on the formula considering radiative and electroweak corrections\cite{ref:Bahcall_1995}. 
Both classes provide the option to use standard formulas without corrections for comparison\cite{ref:tHooft_1971, ref:Bahcall_1987}.

The Monte Carlo simulation of short baseline experiments using radioactive sources is implemented in 
{\tt vModelGenerator} and its subclass, {\tt vModelGeneratorSterile}. The other subclass, {\tt vModelGeneratorSterileEE}, assumes the electron antineutrino initial state and use a simplified formula of oscillation probability to reduce computation time.
The simulation generates the expected spectrum of detected neutrinos at the detector.
The detector and the source are defined with the {\tt vDetector} class and the {\tt vSource} class. Both classes include functions for generating random positions within the detector or the source, simulating the travel distance of the neutrino.
Constants used in the calculations, such as the masses of proton or neutron, are defined in the {\tt vConstant} class.

\begin{figure}[htbp]
\includegraphics[width=\linewidth]{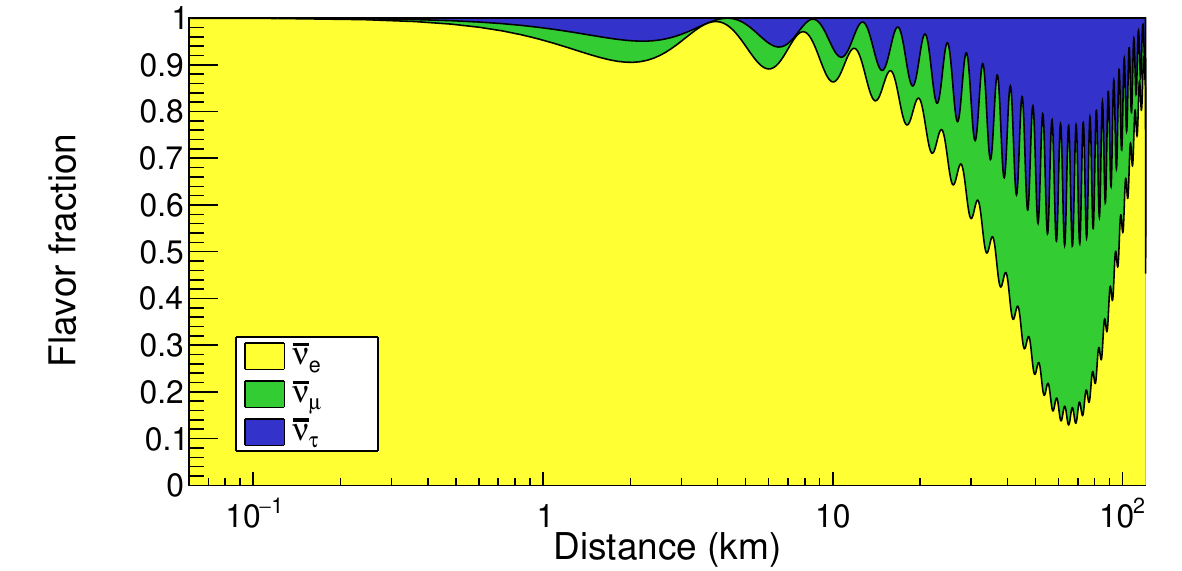}
\caption{
    Fractions of each flavor as a function of distance 
    computed by $\nu${\tt Oscillation},
    are shown 
    when there are only the $\bar\nu_e$ of 4 MeV energy in the initial state. 
    This result is consistent with Fig. 1 of Ref.\cite{ref:nature_JUNO}.
}
\label{fig:validation_oscillation}
\end{figure}

In order to validate the mixing effects implemented in the
package $\nu${\tt Oscillation}, 
we reproduce an illustration of the neutrino oscillation 
in Ref.\cite{ref:nature_JUNO} 
and the result is shown in Fig. \ref{fig:validation_oscillation}. 
It shows the flavor fractions as a function of the propagation distance 
when the neutrino energy is 4 MeV. 
This confirms that our result is at least
qualitatively consistent with the reference by 
comparing the detailed shape of the flavor fraction.

\section{Nuclear Reactor Neutrinos}

\begin{figure}[htbp]
\includegraphics[width=\linewidth]{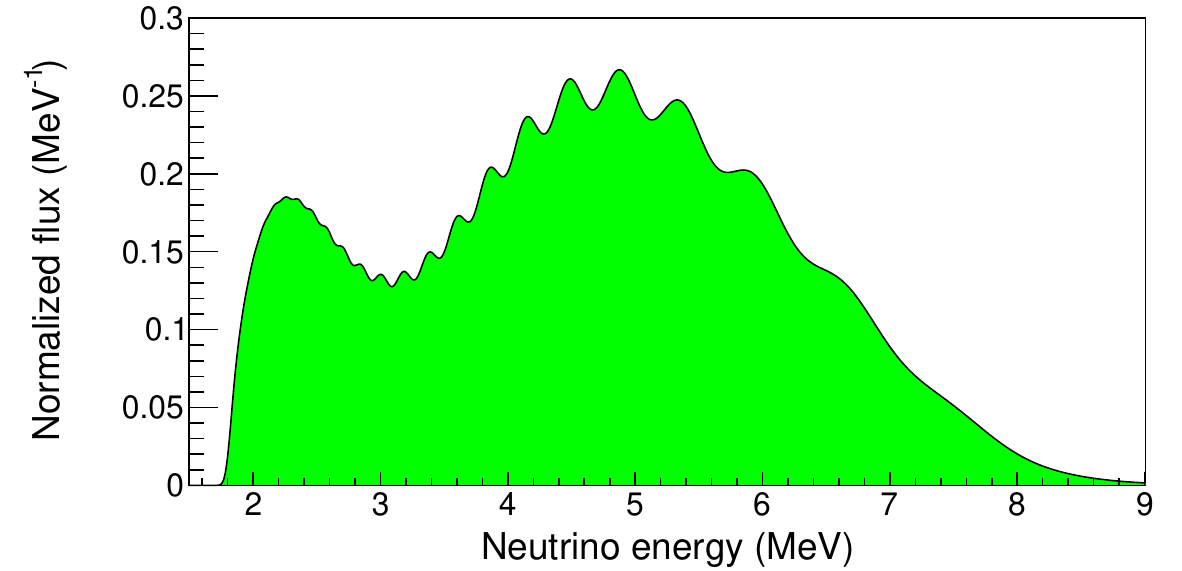}
\caption{
Expected spectrum
of the detected electron-antineutrinos from a reactor
located 53 km away from the detector. The energy resolution of $3\%/\sqrt{E(\text{MeV})}$ is assumed. 
This result is consistent with  
the Fig. 5 of Ref. \cite{ref:nature_JUNO}.
}
\label{fig:validation_reactor}
\end{figure}

The first step in computing the neutrino oscillation
from nuclear reactors is to calculate the flux of anti-neutrinos produced by nuclear fission reactions within the reactors.
For this calculation, we implement two models: the Huber-Mueller\cite{ref:Huber, ref:Mueller} 
and the Gutlein model\cite{ref:Gutlein} in the {\tt vReactorSpectrum} class. 
In the Gutlein model, the isotopes 
$^{235}$U, $^{238}$U, and $^{239}$Pu,
are included, while the Huber-Mueller model includes $^{241}$Pu in addition. This calculation requires input values of the relative abundance of 
all decay isotopes and the thermonuclear 
power (in Watts, we denote $W_\textrm{th}$) 
of a reactor.
Using them as input information, the averaged differential 
neutrino flux 
at a distance $L$ from a nuclear reactor core
is expressed as

\begin{eqnarray}
\Bigg \langle
\frac{d\phi}{dE_{\bar{\nu}_e}}
\Bigg \rangle
= \frac{W_\textrm{th}}{\sum_i f_i E_{\textrm{re},i}}
\frac{1}{4\pi L^2} \sum_i f_i \frac{d \phi_i}{d E_{\bar{\nu}_e}},
\end{eqnarray}
where $W_\textrm{th}$ is the thermal power of the reactor,
$f_i$ is the fraction of a given fission isotope $i$, 
$E_{\textrm{re},i}$ is the thermal energy per fission, and
$d \phi_i/d E_{\bar{\nu}_e}$ is the differential neutrino
flux for the isotope $i$ at the nuclear reactor.
To estimate the detection of antineutrinos from a reactor, one has
to consider two factors. First is the oscillation
effect, and the second is the interaction of the antineutrinos in the detector,
and for that
one has to compute the event rate using the IBD  
cross section. Based on these, the detection probability can be expressed as

\begin{eqnarray}
P^\textrm{det}_{\bar{\nu}_e \rightarrow \bar{\nu}_e} (L)
= \frac{
    \int^\infty_0 \langle 
    \frac{d\phi}{d E_{\bar{\nu}_e}} 
    \rangle \sigma (E_{\bar{\nu}_e}) 
    P_{\bar{\nu}_e \rightarrow \bar{\nu}_e} (L, E_{\bar{\nu}_e} ) 
    d E_{\bar{\nu}_e} 
}
{
\int^\infty_0 \langle 
\frac{d\phi}{d E_{\bar{\nu}_e}} 
\rangle 
\sigma (E_{\bar{\nu}_e}) d E_{\bar{\nu}_e}
},
\end{eqnarray}
where $\sigma(E_{\bar{\nu}_e})$ is the IBD cross section, 
and
$P_{\bar{\nu}_e \rightarrow \bar{\nu}_e} (L)$
is the survival probability for electron anti-neutrino.
Here we validate our computation to the existing nuclear reactor based experiment
by reproducing a result in Fig. 5 of Ref.\cite{ref:nature_JUNO}, and our result is shown in Fig. \ref{fig:validation_reactor}. 
The expected spectrum of the electron-antineutrinos is calculated 
by considering the reactor spectrum, the IBD cross section, and the neutrino oscillation with assuming a reactor located 53 km away from the detector. The energy resolution of $3\%/\sqrt{E(\text{MeV})}$ is assumed.

For a case study with our $\nu${\tt Oscillation},
we choose an underground facility recently constructed in
Yemilab\cite{ref:Yemilab}. 
This is an underground laboratory located at
$37.2^\circ$ N, $128.7^\circ$ E 
in latitude and longitude, and 1000 m below the Yemi mountain of the Republic of Korea. 
We assume a cylindrical detector
with a height and diameter of 15 m, containing 
linear alkylbenzene (LAB) based liquid scintillator
of 2 kton, capable of being installed in a large hall of Yemilab \cite{ref:Yemilab_detector}.
No dopants in the liquid scintillator is assumed in this work.
There are four
nuclear reactors to be considered 
for the reactor anti-neutrino flux 
with nuclear thermal power (name) of 
20.8 (Hanul), 
11.8 (Wolsong), 
21.3 (Gori), and 
16.9 (Hanbit)
$\text{GW}_\text{th}$\cite{ref:IAEApris2022},
and they are 65, 180, 216, and 282 km away from the detector, respectively.

\begin{figure}[htbp]
\includegraphics[width=\linewidth]{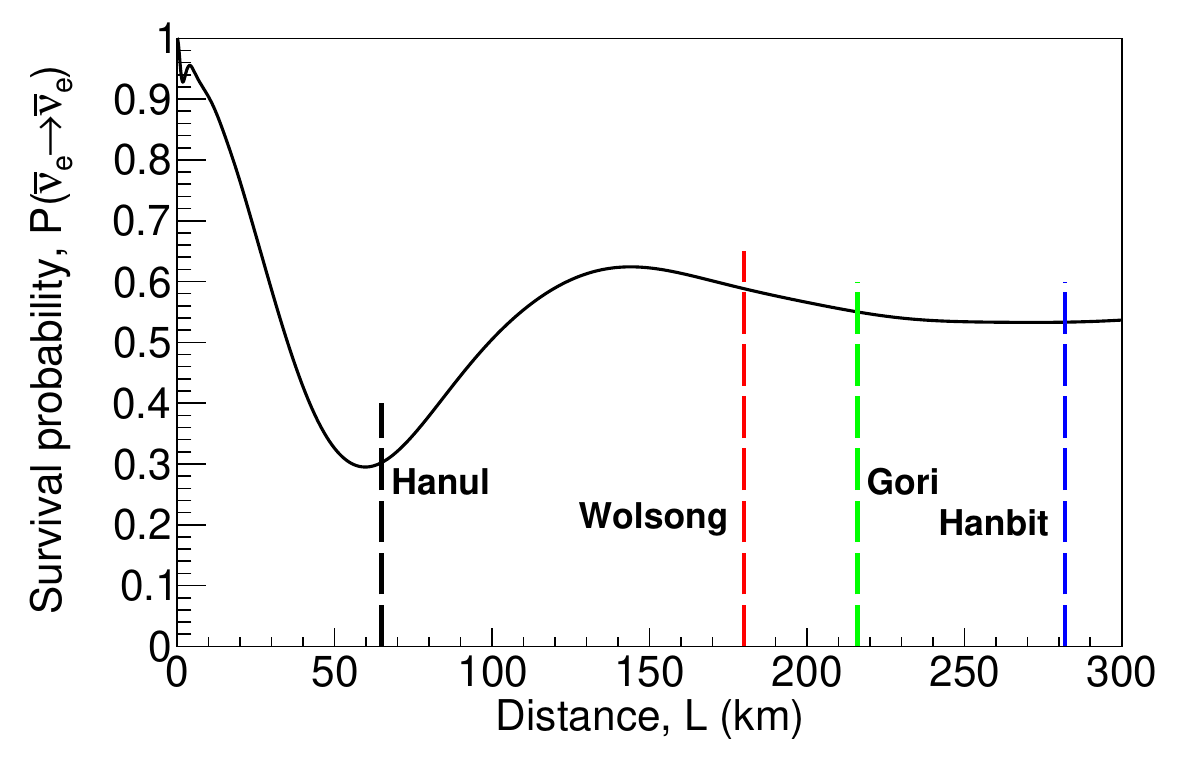}
\includegraphics[width=\linewidth]{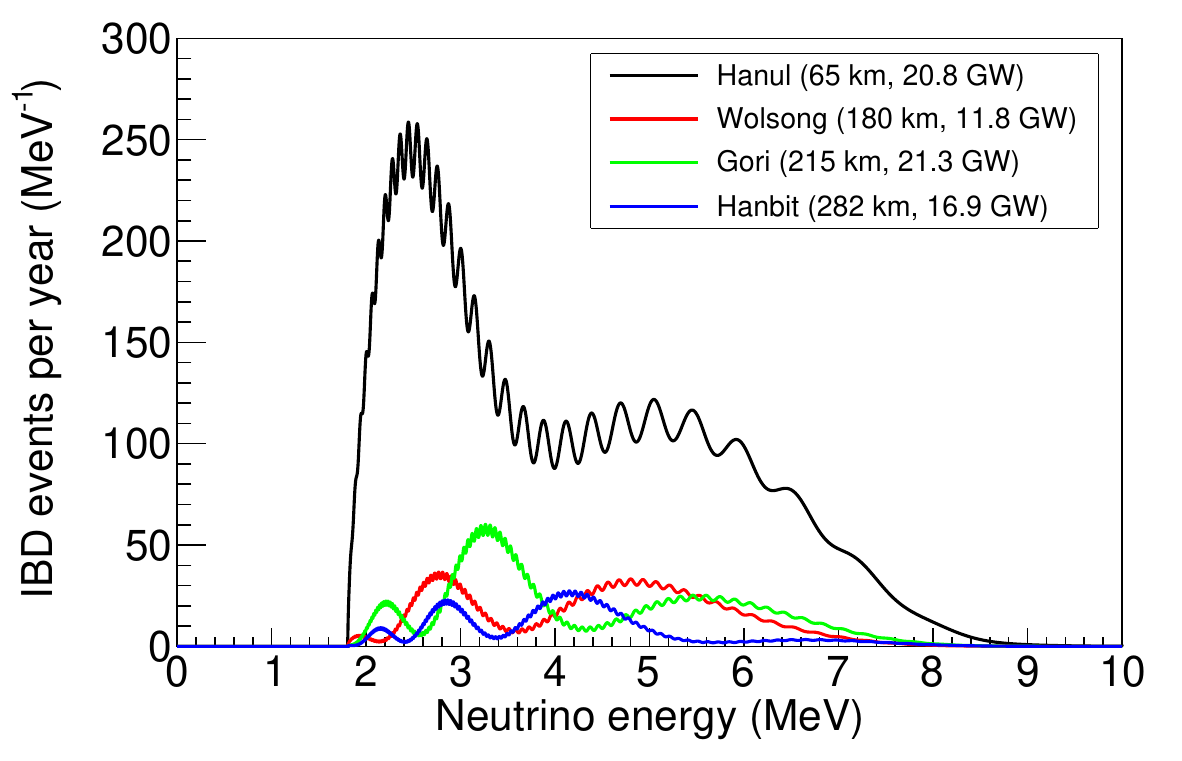}
\caption{
    Top: Our computation of the survival probability
    of anti-electron neutrino as a function of the
    distance from nuclear reactors. 
    The dashed lines indicate the distance between Yemilab and the nuclear power plants: Hanul, Wolsong, Gori, and Hanbit.
    Bottom: Expected spectra of IBD events from the nuclear reactors.
}
\label{fig:reactor_n}
\end{figure}

The distributions of IBD events as a function of distance and neutrino energy
are shown in Fig.~\ref{fig:reactor_n}. 
Despite having the lowest survival probability,
the Hanul reactor is the dominant source of the IBD events as it is closest to the detector site.
The estimated event rate of this source is approximately 860 events per year.

\section{Solar Neutrinos}

The Sun is a natural source of electron neutrinos by nuclear fusion processes
and by now the so-called standard solar model\cite{ref:Bahcall} describes the
flux of electron neutrinos from various different fusion processes very precisely. 
The solar neutrino experiments are measuring the neutrino flux to obtain information of the inner structure of the Sun and the fusion reactions in the Sun.
In our study,
we use parameters from the B16-GS98 model\cite{ref:Vinyoles} to compute the solar neutrino flux. 
Solar neutrinos, arriving at detectors on Earth
interact with detectors via neutrino-electron scattering. 
The differential yields of the scattering at different recoil kinetic energy ($T$)
of the electron can be expressed as

\begin{eqnarray}
\frac{d N_e}{dT} &=& n_e 
\int^{E_{\nu,\textrm{max}}}_{E_{\nu,\textrm{min}}} dE_\nu
\frac{d \Phi}{dE_\nu} (E_\nu)
\Bigg(
\frac{d\sigma_{\nu_e}}{dT} (E_\nu,T) P_{ee} + \frac{d\sigma_{\nu_\mu, \nu_\tau}}{dT} (E_\nu, T)
(1-P_{ee})
\Bigg),
\label{eq:solar}
\end{eqnarray} 
where $n_e$ is the target electron number,
$\frac{d\Phi}{dE_\nu}$ 
is the differential solar neutrino flux,
$P_{ee}$ is the oscillation probability to the electron neutrino final state,
$\frac{d\sigma_{\nu_e}}{dT}$ 
is the differential cross section of 
electron neutrino-electron ($\nu_e$-$e$) scattering\cite{ref:Bahcall_1995}, 
and
$\frac{d\sigma_{\nu_\mu, \nu_\tau}}{dT}$ 
is sum of the differential cross sections of 
$\nu_\mu$-$e$ and $\nu_\tau$-$e$ scattering.

\begin{figure}[htbp]
\includegraphics[width=\linewidth]{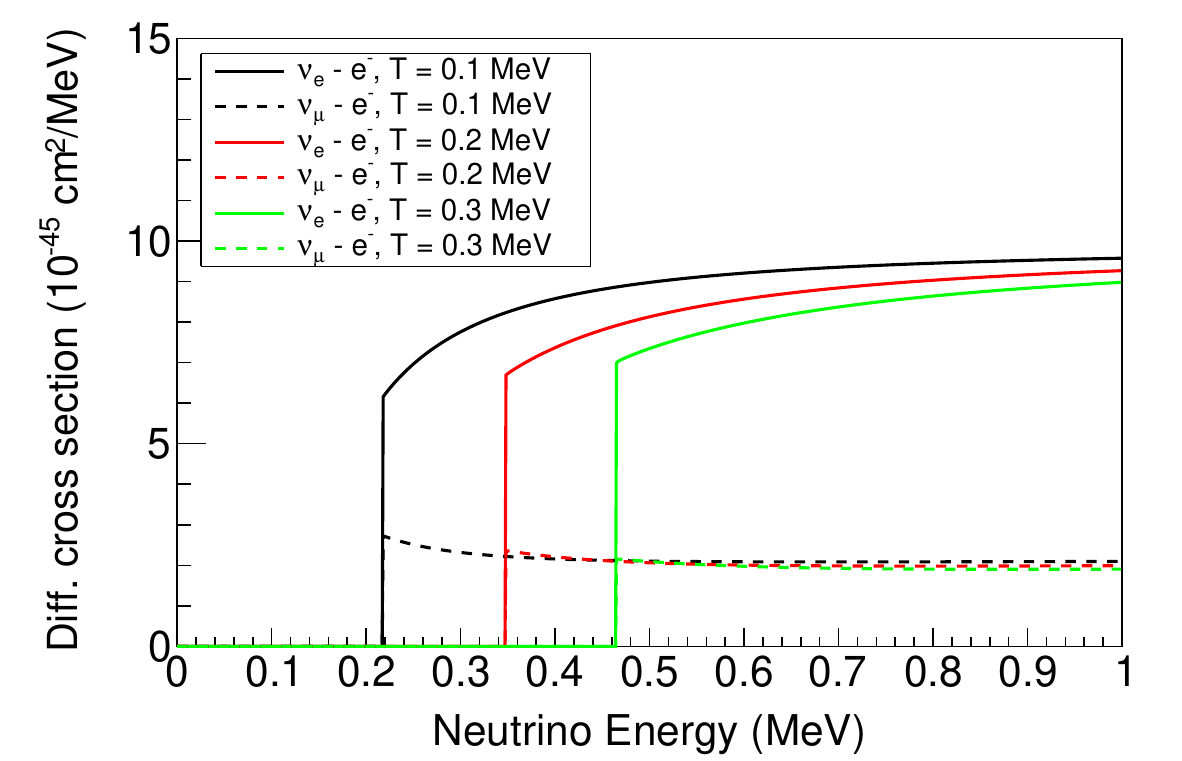}
\caption{
Neutrino-electron elastic scattering
cross sections with different electron recoil energy values.
}
\label{fig:solar}
\end{figure}

\begin{figure}[htbp]
\includegraphics[width=\linewidth]{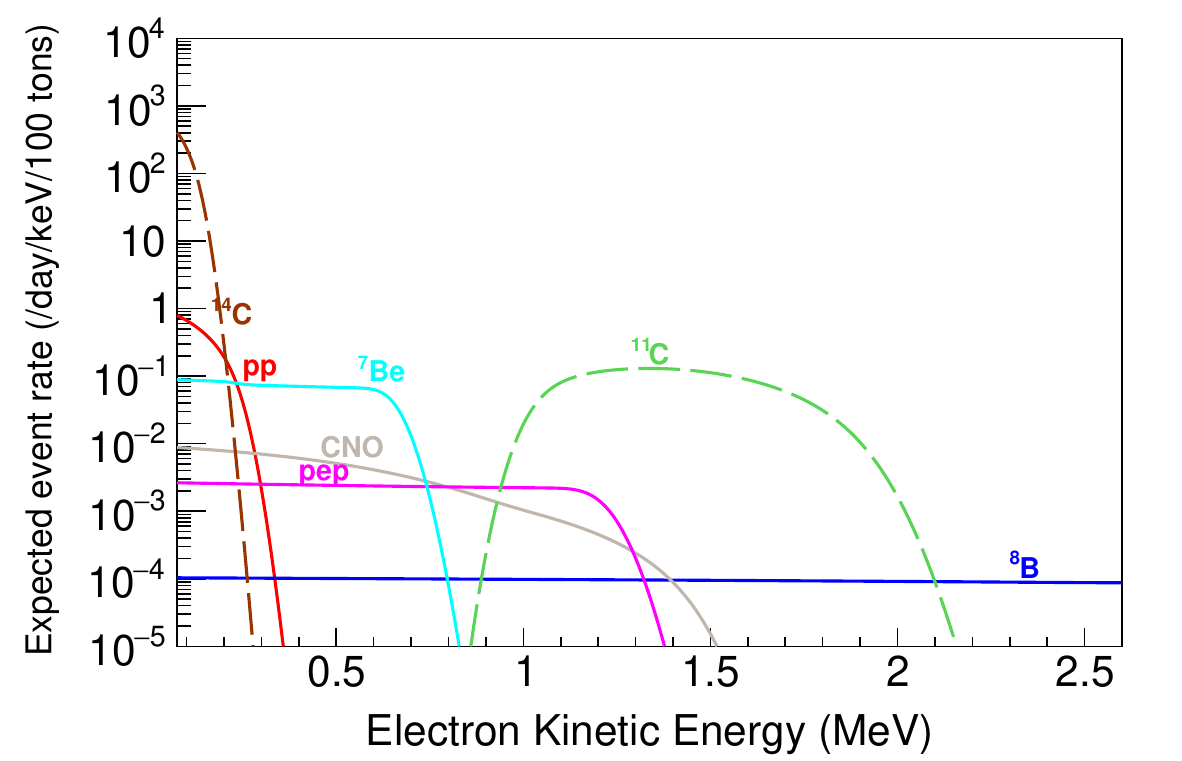}
\caption{
    Expected electron kinetic energy distribution from various fusion processes in the Sun. 
    The yield is normalized to the number per day per 1 keV per 100 tons of a detector.
    Also included two major irreducible backgrounds from $^{14}\textrm{C}$ and $^{11}\textrm{C}$. 
}
\label{fig:solar_rate}
\end{figure}

Figure~\ref{fig:solar} shows the implemented
cross sections 
of neutrino-electron scatterings 
in Eq.~(\ref{eq:solar}) at different electron
recoil energy values with different 
flavors of neutrinos. They are integrated
with the solar neutrino flux model mentioned
earlier in order to obtain spectra
of scattered electron kinetic energy. 
Such spectra
for $pp$, $pep$, CNO, $^7\textrm{Be}$,
and $^{8}\textrm{B}$ neutrinos are shown in 
Fig.~\ref{fig:solar_rate}. 
The yield is normalized
to the number of events per day per 1 keV per 100 ton of a 
liquid scintillator and the energy resolution of $5\%/\sqrt{E(\text{MeV})}$ is assumed. 
The normalization and shapes are in general
in agreement with results in Ref.\cite{ref:Borexino2021}.
We also include two irreducible backgrounds.
The first one is $^{14}\textrm{C}$ background. $^{14}\textrm{C}$ is a natural isotope which emits electron antineutrino through beta decay to $^{14}\textrm{N}$ with decay energy of 0.156 MeV and half life of 5730 years\cite{ref:KAERI_C14}.
This background is
mainly populated below 0.3 MeV.
Here we assume the concentration of 
$^{14}\text{C}/^{12}\text{C} = 1.94\times10^{-18}$\cite{ref:Borexino1998}. 
The second one is the so-called cosmogenic
$^{11}\textrm{C}$ background.
$^{11}\textrm{C}$, produced in a reaction induced by cosmic muons, emits positron and electron neutrino through positive beta decay with decay energy of 0.96 MeV and half life of 20 minutes. 
Because the energy deposition of the positron and the release of the two gammas from electron-positron annihilation are close to the same time,
the visible energy is in the range of 0.8 MeV and 2.2 MeV \cite{ref:Borexino2021}.
$^{11}\textrm{C}$ is produced by residual cosmic muons and is therefore sensitive to the depth of the experimental site. Our assumed site (or Yemilab) has a 1000 m overburden and the Borexino detector is installed at a depth of 1.4 times that, so it shows a different event rate with the reference spectrum in Ref. [32].
In addition, the range over which the 
$^{11}\textrm{C}$ background spectrum is distributed is different from the reference spectrum. This is because our results do not include detector responses such as quenching as they are detector specific information.

\section{Sterile Neutrinos}

Recently, several experiments indicate that the survival 
probability of electron antineutrinos in short range may be smaller than expected\cite{ref:LSND_anomaly, ref:miniboone_anomaly, ref:stereo_anomaly}. 
This may originate due to the presence of a different type of neutrino, so-called
sterile neutrino. We implement a (3+1) neutrino model in the $\nu${\tt Oscillation}, 
in which four mass eigenstates
including one sterile neutrino, are mixed to construct three flavors. 

\begin{figure}[htbp]
\includegraphics[width=\linewidth]{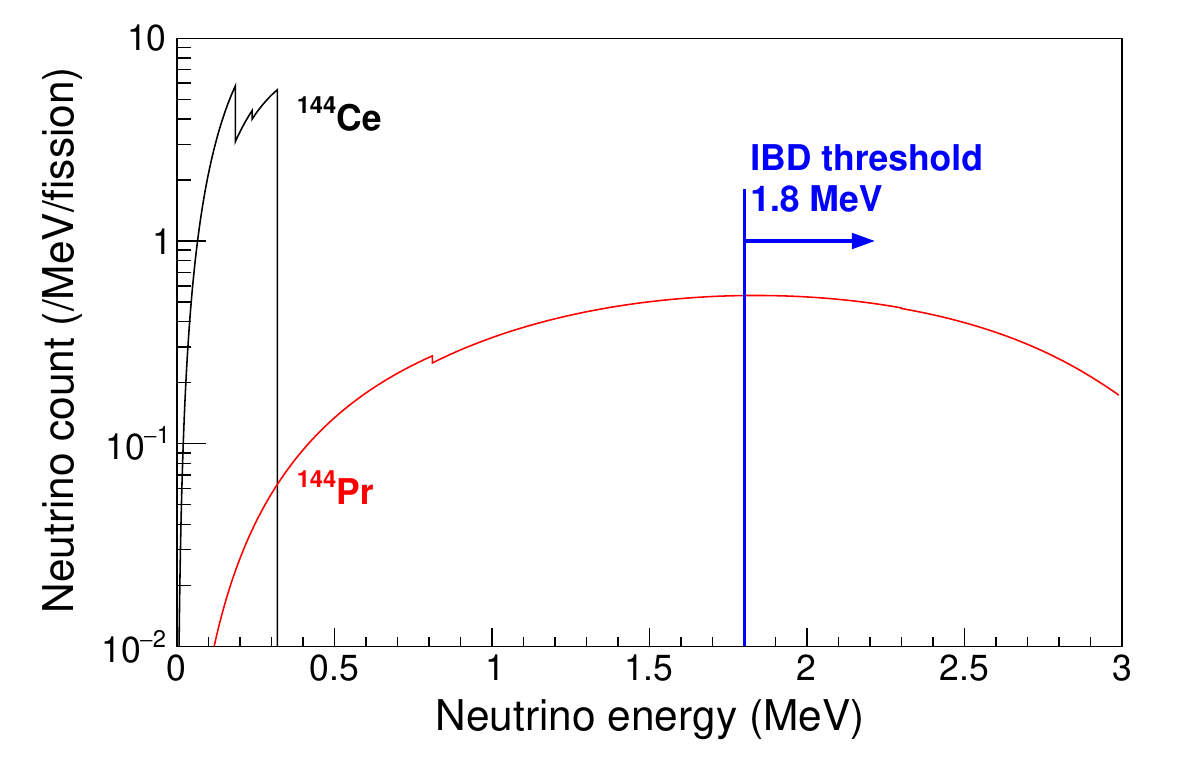}
\caption{
Energy distributions of neutrinos from $\beta^-$ decays
of $^{144}\textrm{Ce}$ and $^{144}\textrm{Pr}$. 
}
\label{fig:ce144}
\end{figure}

In order to
estimate the sensitivity to the sterile neutrino search with the given
radioactive source, we assume a radioactive source of $^{144}$Ce with activity of 100 kCi as proposed earlier by other experiments\cite{ref:celand, ref:dayabay, ref:sox}.
We again assume the same detector which is presented in the reactor neutrino section.
The detector hall at Yemilab is connected by two tunnels, one of which is located at the mid-height of the detector. 
The source is assumed to be located in the tunnel at a distance of 9.5 m from the center of the detector.
The $^{144}\textrm{Ce}$ decays to $^{144}\textrm{Pr}$ with the half life of 285 days, 
and $^{144}\textrm{Pr}$ decays to $^{144}\textrm{Nd}$ with the half life of 17 minutes, emitting $\bar{\nu}_e$. The energy distributions
of anti-neutrinos from radioactive sources are shown in Fig.~\ref{fig:ce144}. Note that,
since the threshold for IBD is 1.8 MeV, 
IBD is sensitive only to neutrinos from $^{144}\textrm{Pr}$.
These spectra are implemented in {\tt vCe144Spectrum} class accordingly.

\begin{figure}[htbp]
\includegraphics[width=\linewidth]{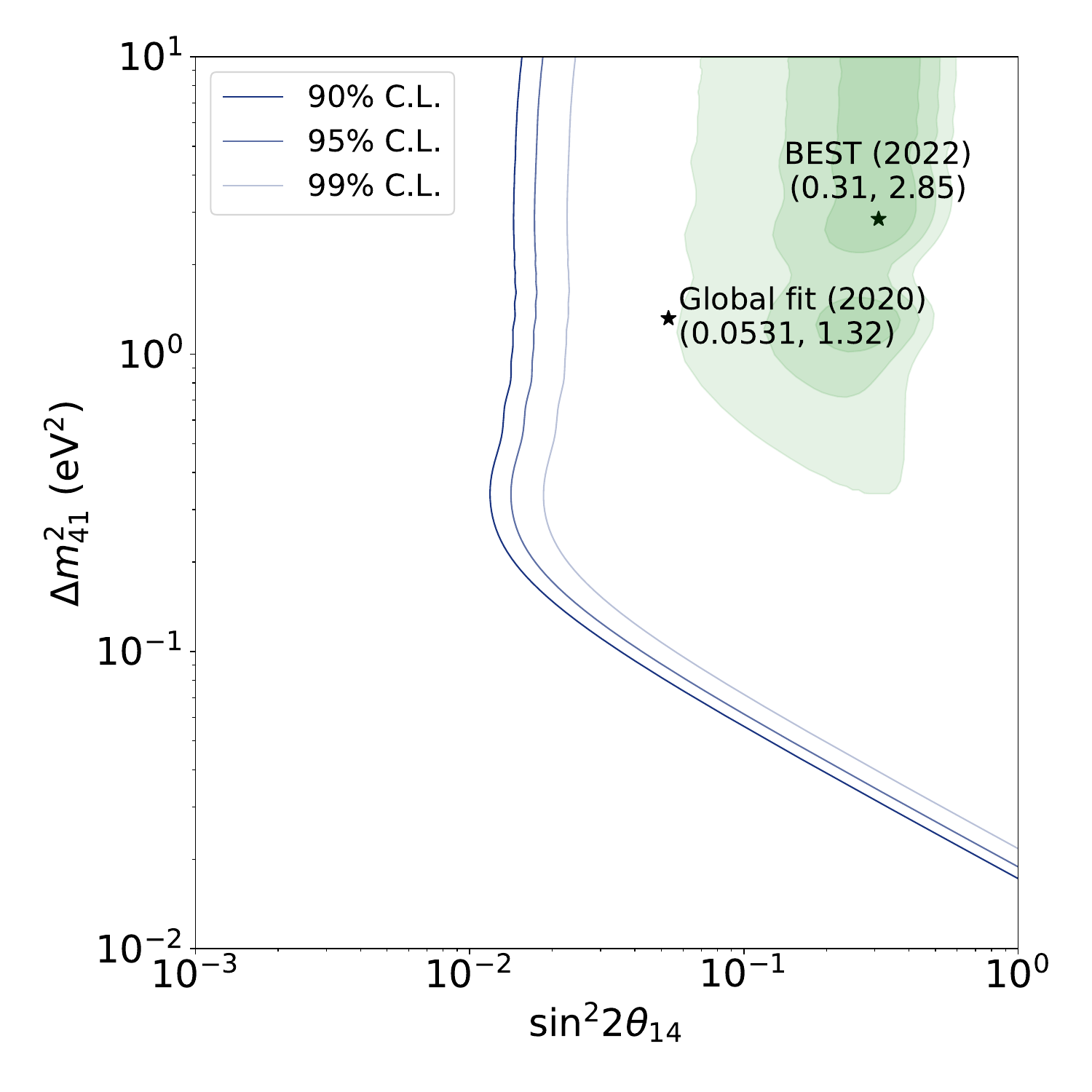}
\caption{
Exclusion sensitivity of the (3+1) neutrino model at 90\%, 95\%, and 99\% confidence levels. 
A simulated data set of 34,000 IBD events without including sterile neutrino is compared with the (3+1) models. 
The one year of data taking and 70\% detection efficiency are assumed. The best fit of global analysis of disappearance data and the BEST experiment are also marked, along with the $1$-$3\sigma$ confidence levels of the result of BEST indicated with green contours.
}
\label{fig:sterile_lk_contour}
\end{figure}

We simulate a data set without including the sterile neutrino oscillations.
In total 34,000 neutrino events are generated by assuming the one year of data taking with 70\% detection efficiency\cite{ref:JUNO_efficiency}. The resolution for energy and position are not considered in this data. 
By comparing this data to the expected models with the sterile neutrino oscillations, 
we quantify the exclusion sensitivity at 90\%, 95\% and 99\%
confidence levels as shown in Fig. \ref{fig:sterile_lk_contour}. 
The best fit of the global analysis of disappearance data\cite{ref:globalfit2020} and the result of the Baksan Experiment on Sterile Transitions (BEST)\cite{ref:best2022} are also marked. This analysis is intended to check the feasibility of this package, and the result and assumptions may be optimistic. We plan to conduct detailed analysis to estimate the sensitivity of the detector.

\section{Conclusions and Outlook}

We developed a software package, $\nu${\tt Oscillation},
that is capable to calculate and simulate 
the production, oscillation, and detection of neutrinos.
We validated the package by reproducing the simulations on the neutrino spectra of reactor, solar and source neutrinos and the detection of them.
We also assessed the feasibility of this package by estimating the sensitivity of a liquid scintillation detector to the sterile neutrino. 
The current version of this package does not include the matter effect of the earth, which becomes significant in long-base line setups. This feature will be implemented in next releases to extend the scope of application.
This work will be used for the design and optimization of our experiment and is expected to benefit future experiments.

\begin{acknowledgments}

This work was supported 
by the National Research Foundation of Korea (NRF) grant funded
by the Korea government (MSIT) (No. NRF-2022R1A2B5B02001535) 
and Institute for Basic Science (IBS) under the project code IBS-R016-D1.

\end{acknowledgments}

\appendix

\section{An example how to use the package}

In this section, we demonstrate how to use the package by using the calculation for Fig. 3. In this calculation, the energy spectrum of electron anti-neutrino from a reactor, survival probability of the neutrino after oscillation, and the IBD cross section regarding the detector. The parameters to be used are taken from JUNO experiment and its surrounding reactors. 
The software package includes several other examples, providing users with guidance on using the package.

The following scripts are compatible with ROOT framework.
First, include the software package in the script:
\begin{verbatim}
    #include "vClass.hh"
\end{verbatim}
Alternatively, the required headers or scripts can be included individually:
\begin{verbatim}
    #include "vReactorSpectrum.cc" 
    #include "vOscillation.cc" 
    #include "vIBD.cc" 
    #include "vConstant.hh" 
\end{verbatim}
Next, declare the objects required for the calculation. 
\begin{verbatim}
    vReactorSpectrum *vRe = new vReactorSpectrum("HM");
    vIBD *ibd = new vIBD();
    vOscillation *vosc = new vOscillation();
\end{verbatim}
In this example, the Huber-Mueller (HM) model is utilized for the reactor neutrino spectrum calculation. The oscillation parameters are set to default values, but they can be modified using the following functions:
\begin{verbatim}
    vosc->Settheta(theta_12, theta_13, theta_23);
    vosc->SetDm2(Dm2_21, Dm2_31);
    vosc->Setdelta_CP(delta_CP);
\end{verbatim}
where {\tt Settheta}, {\tt SetDm2}, and {\tt Setdelta\_CP} are functions to set the mixing angles, mass-squared differences, and CP phase, respectively.
The following lines compute the flux of reactor neutrinos for a given neutrino energy {\tt E},
\begin{verbatim}
    double Nreactor = vRe->GetCurrent(E);
\end{verbatim}
the IBD cross section at a given neutrino energy {\tt E},
\begin{verbatim}
    double ibdcs = ibd->GetCrossSection_integral(E);
\end{verbatim}
and the oscillation probability for a given distance {\tt L}, energy {\tt E}, and initial and final state of electron neutrinos.
\begin{verbatim}
    double Pee = vosc->GetProbability(L, E, "e", "e");
\end{verbatim}
Then the expected probability of detecting an electron anti-neutrino from the reactor for the given energy is obtained by multiplying above three results:
\begin{verbatim}
    double Pdet = Nreactor * ibdcs * Pee;
\end{verbatim}
To complete Figure 3, we compute the {\tt Pdet} values over the useful energy range. Then we apply it normalization and convolution with a Gaussian function to account the $3\%/\sqrt{E(\text{MeV})}$ energy resolution.

\end{document}